\RequirePackage{fix-cm}
\documentclass[smallextended]{svjour3}       
\smartqed  

\usepackage{epic}
\usepackage{latexsym}
\usepackage{amsmath}
\usepackage{epsfig}
\usepackage{amssymb}
\usepackage{enumerate}
\usepackage{color}
\usepackage{mathptmx}     
\usepackage{subcaption}
\usepackage{natbib}
\DeclareRobustCommand{\VAN}[3]{#2}

\usepackage{graphicx}
\usepackage{enumerate}
\usepackage{color}
\newcommand{\blue}{\textcolor{black}}
\newcommand{\red}{\textcolor{black}}

\journalname{Journal of Mathematical Biology}
\begin{document}

\title{\blue{Tree-based} networks: characterisations, metrics, and support trees
\thanks{[JCP] Partially supported by the Spanish Ministry of Economy and
Competitiveness and European Regional Development Fund project
DPI2015-67082-P (MINECO/FEDER)}
}
%



\author{Joan Carles Pons        \and Charles Semple \and Mike Steel
}


\institute{J.C. Pons \at
  Department of Mathematics and Computer Science, 
  University of the Balearic Islands, Palma, Spain\\
  \email{joancarles.pons@uib.es}           
            \and
          C. Semple, M. Steel \at
School of Mathematics and Statistics, University of Canterbury, 
 Christchurch, New Zealand \\
  \email{\{charles.semple, mike.steel\}@canterbury.ac.nz} 
}

\date{Received: date / Accepted: date}

\maketitle

\begin{abstract}
Phylogenetic networks generalise phylogenetic trees and allow for the accurate representation of the evolutionary history of a set of present-day species whose past includes reticulate events such as hybridisation and lateral gene transfer. One way to obtain such a network is by starting with a (rooted) phylogenetic tree $T$, called a base tree, and adding arcs between arcs of $T$. The class of phylogenetic networks that can be obtained in this way is called tree-based networks and includes the prominent classes of tree-child and reticulation-visible networks. Initially defined for binary phylogenetic networks, tree-based networks naturally extend to arbitrary phylogenetic networks. In this paper, we generalise recent tree-based characterisations and associated proximity measures for binary phylogenetic networks to arbitrary phylogenetic networks. These characterisations are in terms of matchings in bipartite graphs, path partitions, and antichains. Some of the generalisations are straightforward to establish using the original approach, while others require a very different approach. Furthermore, for an arbitrary tree-based network $N$, we characterise the support trees of $N$, that is, the tree-based embeddings of $N$. We use this characterisation to give an explicit formula for the number of support trees of $N$ when $N$ is binary. This formula is written in terms of the components of a bipartite graph.

\end{abstract}

\section{Introduction}

Classically, the evolution of species has been assumed to be a branching process represented by a phylogenetic (evolutionary) tree. However, reticulate evolutionary processes such as hybridisation (e.g., in plants and some groups of animals), endosymbiosis (in early life), and horizontal gene transfer (e.g., in bacteria) can only be represented appropriately by using phylogenetic networks~\citep{doolittle2007pattern}.

One way to represent these reticulate processes is to take a rooted binary phylogenetic tree $T$, called a base tree, and sequentially add arcs by subdividing branches of $T$. Of course, not all phylogenetic networks can be obtained in this way~\citep{van2013different}. However, those phylogenetic networks that can arise in this way form a rich class of networks called tree-based networks~\citep{francis2015phylogenetic} and include the well-studied classes of tree-child~\citep{cardona2009comparison}, tree-sibling~\citep{cardona2008distance}, and reticulation-visible~\citep{huson2010} networks. The biological emphasis of this distinction is to determine whether the evolution of some groups is mainly tree-like with reticulations between the branches, or whether the reticulation is so entangled that no tree-like description is reasonable \citep{dagan2006tree, doolittle2007pattern}. As well as the viewpoints emphasised in this paper, tree-based networks have been studied in a variety of ways. For example, see \cite{anaya2016determining}, \cite{francis2017tree}, \cite{hayamizu}, \cite{semple2016phylogenetic}, and \cite{zhang2016Zgraph}.

Binary tree-based networks can be mathematically characterised in a number of different ways.  Some of these characterisations are in terms of bipartite matchings~\citep{zhang2016Zgraph, van2016nonbinary}. Others, are based on antichains and path partitions~\citep{francis2016new}. These characterisations have allowed the development of computationally efficient indices to measure how `close'  an arbitrary binary phylogenetic network is to being tree-based \citep{francis2016new}.

In applied phylogenetics,  binary phylogenetic trees and networks are often overly restrictive  \citep{morrison2011introduction}.  One reason for this is that vertices of out-degree greater than 2 (called `polytomies' in biology) are used to represent uncertainty about the precise order of speciation events (a `soft' polytomy), or to indicate rapid species radiation, such as might occur with the arrival of a species on a new island leading to the near-simultaneous evolution of several new species  (a `hard' polytomy). Also, some reticulation events may best be represented by vertices with more than two parents simultaneously (for example, when  two binary hybridisation or reticulation events are separated by a very short period of time and where the order of the events may not be clear). 
In summary, divergence and reticulation events result in phylogenetic networks in which some (possibly all) vertices have either in-degree or out-degree greater than two, that is, nonbinary phylogenetic networks.

The notion of tree-based (described informally in the second paragraph above) was extended from binary  to non-binary networks in ~\cite{van2016nonbinary}, by allowing  the  addition of arcs between branches and vertices of a base tree (a more precise definition is given  in the next section). 
\cite{van2016nonbinary} also defined and investigated  a more restrictive class of phylogenetic networks, referred to as `strictly tree-based', which allow the addition of arcs only between branches of the base tree, and with no two of the additional arcs attaching to the same vertex (thus the nonbinary nature of the network arises purely from the nonbinary nature of the base tree). We will not consider this more restricted notion further in this paper.

In the first part of the paper (Section~\ref{part1}), we generalise the tree-based characterisations of binary phylogenetic networks and their derived deviation indices given in \cite{francis2016new} to arbitrary phylogenetic networks.  If $N$ is a tree-based network (not necessarily binary), a support tree of $N$ is an embedding of a base tree for $N$. In the second part of the paper (Section~\ref{part2}), we consider the problem of counting the number of support trees of $N$. To this end, we introduce a bipartite graph associated with $N$ that characterises the support trees of $N$. This characterisation is used to determine this number when $N$ is binary. Moreover, we show that this determined number equates to the upper bound of the number of base trees of a binary tree-based network established in~\cite{jetten2015characterising}. \blue{Counting the number of support trees in the non-binary setting is left for future work.}

\section{Preliminaries}

Throughout the paper, $X$ denotes a non-empty finite set. Given a set $X$ of taxa, a (rooted) \emph{phylogenetic network on $X$} is a rooted acyclic digraph $N$ with no parallel arcs satisfying the following properties:
\begin{enumerate}[(i)]
\item the unique root has out-degree at least one;

\item the set $X$ is the set of vertices of out-degree zero, each of which has in-degree one;

\item all other vertices either have in-degree one and out-degree at least two, or in-degree at least two and out-degree one.
\end{enumerate}
If $|X|=1$, we additionally allow $N$ to consist of the single vertex in $X$. The vertices in $X$ are called {\em leaves}, while the vertices of in-degree one and out-degree at least two are {\em tree vertices} and the vertices of in-degree at least two and out-degree one are {\em reticulations}. Furthermore, a vertex in $N$ is an \emph{omnian} if it is a non-leaf vertex whose children are all reticulations. An arc ending in a reticulation is a \emph{reticulation arc}; all other arcs are {\em tree arcs}. We say $N$ is a {\em binary} phylogenetic network if each tree vertex has out-degree two and each reticulation has in-degree two.

A (rooted) \emph{phylogenetic $X$-tree} is a phylogenetic network on $X$ that contains no reticulations. Thus a {\em binary} phylogenetic $X$-tree is a phylogenetic $X$-tree in which the root has out-degree either one or two, and all other non-leaf vertices have in-degree one and out-degree two.

For an arbitrary directed graph $D$, {\em subdividing} an arc $(u, v)$ in $D$ is the operation of replacing $(u, v)$ by two arcs $(u, w)$ and $(w, v)$, where $w$ is a new vertex. A {\em subdivision} of $D$ is a directed graph obtained from $D$ by a sequence of arc subdivisions. Conversely, {\em suppressing} a vertex $w$ of in-degree one and out-degree one in $D$ is the operation of replacing the two arcs, $(u, w)$ and $(w, v)$ say, incident with $w$ by the single arc $(u, v)$, and finally delete the vertex $w$.

A phylogenetic $X$-tree $T$ is \emph{embedded} in a phylogenetic network $N$ on $X$ if a subdivision of $T$ can be obtained from $N$ by deleting arcs and degree-one vertices. If $T$ can be embedded in $N$, we say that $N$ {\em displays} $T$.

\subsection{Tree-based networks}

\begin{figure}[ht]
\centering
\includegraphics[scale=1.0]{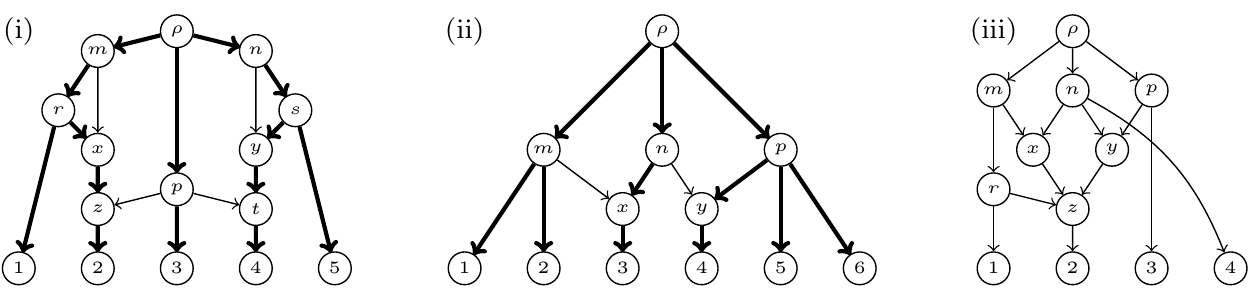}
 \caption{(i) A binary tree-based network and (ii) a nonbinary tree-based network, where a base tree for each network is indicated in bold (the other edges are linking arcs). (iii) A phylogenetic network that is not tree-based.}
 \label{figure1}
\end{figure}

Let $T$ be a phylogenetic $X$-tree. Following~\cite{francis2015phylogenetic} and~\cite{van2016nonbinary}, a phylogenetic network $N$ on $X$ is \emph{tree-based} with {\em base tree} $T$ if $N$ can be obtained from $T$ in the following way. First, add new vertices, called {\em attachment points}, by taking a subdivision of $T$, and then add new arcs $(u, v)$, where either $u$ and $v$ are both attachment points, or $u$ is a non-leaf vertex in $T$ and $v$ is an attachment point. These additional arcs are referred to as {\em linking arcs}. The subdivision of $T$ is called a {\em support tree} for $N$. Note that the set of vertices of a support tree for $N$ is also the set of vertices of $N$. Also, observe that even if $T$ is binary, this construction can lead to a tree-based network in which reticulations have more than two parents and tree vertices have more than two children. To illustrate, Fig.~\ref{figure1} shows three phylogenetic networks. The first two networks are tree-based, while the third network is not tree-based.

Given two phylogenetic networks $N$ and $N'$ on $X$, we say that $N'$ is a \emph{binary refinement} of $N$ if $N$ can be obtained from $N'$ by a sequence of arc contractions. The next lemma describes two alternative, but equivalent, ways of viewing the notion of being `tree-based'. The equivalence of (i) and (ii) was noted in \cite{van2016nonbinary}, while the equivalence of (i) and (iii) is immediate from the definition.

\begin{lemma}
\label{lemeqz}
Let $N$ be a phylogenetic network on $X$. Then the following are equivalent:
\begin{enumerate}[{\rm (i)}]
\item $N$ is tree-based.

\item There exists a binary refinement $N'$ of $N$ that is tree-based.

\item $N$ has a rooted spanning tree with the same root as $N$ and all its leaves in $X$.
\end{enumerate}
\end{lemma}

\begin{figure}[ht]
\centering
\includegraphics[scale=1.0]{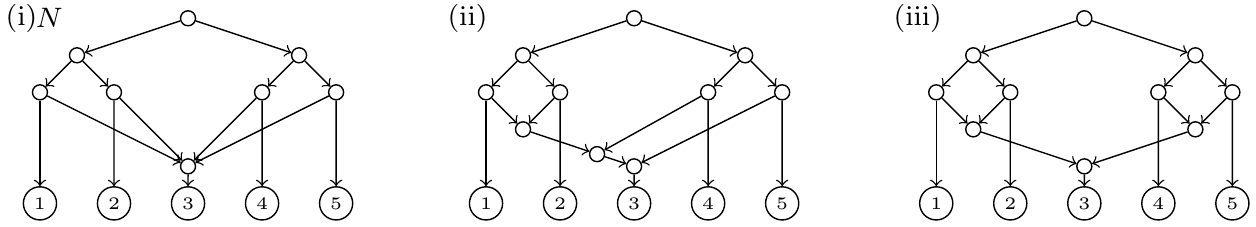}
 \caption{(i) A tree-based network $N$. (ii) A binary refinement of $N$ that is tree-based. (iii) A binary refinement of $N$ that is not tree-based.}
 \label{figure2}
\end{figure}

Although Lemma~\ref{lemeqz} says at least one binary refinement of a tree-based network $N$ is tree-based, it is possible that there is a binary refinement $N'$ of $N$ that is not tree-based. An example to illustrate this is provided by the tree-based network $N$ in Fig.~\ref{figure2}(i), which has (ii) a binary refinement that is tree-based and (iii) another that is not.

\section{Characterisations and metrics}
\label{part1}

\subsection{Bipartite graphs and tree-based characterisations}
\label{sec:bip_tree}

Tree-based networks have been characterised in a variety of ways, particularly in the binary setting. Here we describe two of those characterisations both in terms of matchings in bipartite graphs. The first, due to \cite{zhang2016Zgraph}, is restricted to the binary setting.

Let $N=(V, A)$ be a binary phylogenetic network. Let $T_N$ be the set of tree vertices of $N$ whose children contain at least one reticulation, and let $R$ be the set of reticulations of $N$. Let $\mathcal Z_N$ be the bipartite graph with vertex bipartition $\{T_N, R\}$ and arc set
$$\{(t, r): t\in T_N, r\in R, (t, r)\in A\}.$$
\cite{zhang2016Zgraph} established the following characterisation.

\begin{theorem}
\label{teo:zhang} 
Let $N$ be a binary phylogenetic network. Then the following are equivalent:
\begin{enumerate}[{\rm (i)}]
\item $N$ is tree-based.

\item There is a matching $M$ in $\mathcal{Z}_N$ with $|M|=|R|$. 

\item $\mathcal{Z}_N$ has no maximal path that starts and ends with reticulations.
\end{enumerate}
\end{theorem}

\noindent As we shall see below, Theorem~\ref{teo:zhang} does not extend to arbitrary phylogenetic networks. However, there is an analogous characterisation for arbitrary tree-based networks due to \cite{van2016nonbinary}.

Recall that an omnian is a non-leaf vertex whose children are all reticulations. For an arbitrary phylogenetic network $N=(V, A)$, let $O$ denote the set of omnians of $N$ and let $R$ denote the set of reticulations of $N$. Note that these two sets are not necessarily disjoint, that is, an omnian can also be a reticulation. Let $\mathcal{B}_N$ be the bipartite graph with vertex bipartition $\{O, R\}$ and arc set
$$\{(o, r): o\in O, r\in R, (o, r)\in A\}.$$
\cite{van2016nonbinary} proved the following theorem.

\begin{theorem}
\label{teo:iersel_charact}
Let $N$ be a phylogenetic network. Then the following are equivalent:
\begin{enumerate}[{\rm (i)}]
\item $N$ is tree-based.

\item There is a matching $M$ in $\mathcal{B}_N$ with $|M|=|O|$.
\end{enumerate}
\noindent Moreover, the condition
\begin{enumerate}[{\rm (i)}]
\item[{\rm (iii)}] $\mathcal{B}_N$ has no maximal path that starts and ends with omnians
\end{enumerate}
implies (i) and (ii) and, if $N$ is binary, it is equivalent to each of (i) and (ii).
\end{theorem}

\noindent Theorem~3.4 in \cite{jetten2015characterising} shows that (iii) implies (i), but Fig.~16 in \cite{van2016nonbinary} shows that (i) does not imply (iii) in general.

\begin{figure}[ht]
\centering
 \includegraphics[scale=1.0]{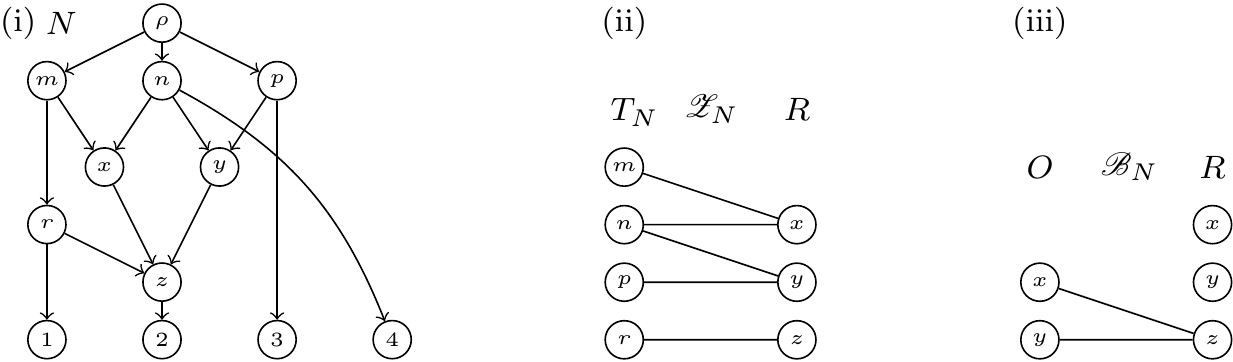}
 \caption{(i) The phylogenetic network $N$ that is not tree-based, reproduced from Fig.~\ref{figure1}(iii); (ii) its associated bipartite graph $\mathcal{Z}_N$; (iii) its associated bipartite graph $\mathcal{B}_N$.
 }
 \label{fig:net2}
\end{figure}

To see that Theorem~\ref{teo:zhang} does not extend directly to arbitrary phylogenetic networks, consider the phylogenetic networks shown in Fig.~\ref{figure1} and Fig.~\ref{fig:net2}. The phylogenetic network $N_1$ shown in Fig.~\ref{figure1}(i) is a tree-based network. However, $\mathcal{Z}_{N_1}$ has  two reticulate vertices $z$ and $t$ that
each have just one tree-vertex parent, namely $p$, and so $\mathcal{Z}_{N_1}$ has no matching that covers each reticulation.
Moreover, by Theorem~\ref{teo:iersel_charact}, the phylogenetic network $N$ shown in Fig.~\ref{fig:net2}(i) is not tree-based as there does not exist a matching in $\mathcal{B}_{N}$ that covers $O$ (see Fig.~\ref{fig:net2}(iii)).  However, $\mathcal{Z}_{N}$, shown in Fig.~\ref{fig:net2}(ii), has a matching in which each reticulation is matched, and $\mathcal{Z}_{N}$ has no maximal path that starts and ends with reticulations.

\subsection{Generalising tree-based characterisations}

In this subsection, we show that many of the characterisations of binary tree-based networks in~\cite{francis2016new} can be extended to arbitrary tree-based networks. These characterisations are in terms of antichains and path partitions as well as matchings in bipartite graphs. Several of the characterisations strengthen the so-called antichain-to-leaf property which is a necessary, but not sufficient, condition for a phylogenetic network to be tree-based (\cite{francis2015phylogenetic}). An {\em antichain} in a directed graph is a subset $S$ of vertices with the property that, for all distinct $u, v\in S$, there is no directed path from $u$ to $v$. A phylogenetic network $N$ satisfies the {\em antichain-to-leaf property} if, for every antichain of $k$ vertices, there exists $k$ vertex-disjoint paths from the elements of the antichain to the leaves of $N$. The binary phylogenetic network shown in Fig.~3 of~\cite{francis2016new} satisfies the antichain-to-leaf property, but it is not tree-based.

To formally state these characterisations, let $D=(V, A)$ be any directed graph. We denote by $\mathcal G_D$ the bipartite graph whose vertex bipartition is $\{V_1, V_2\}$, where each of $V_1$ and $V_2$ is a copy of $V$, and whose arc set is
$$\{(u, v): u\in V_1, v\in V_2, (u, v)\in A\}.$$
This bipartite graph has been referred to as the `bipartite representation' of $D$ \citep{jensen}. For when $D$ is a binary phylogenetic network, $\mathcal G_D$ played a key role in~\cite{francis2016new}.

The next theorem extends the characterisations of binary tree-based networks of Theorem~2.1 in~\cite{francis2016new} to arbitrary tree-based networks.   Its proof consists of straightforward modifications to the proof of Theorem~2.1 in~\cite{francis2016new}, and so it is in the Appendix. As noted above, in the general setting the graph  $\mathcal Z_N$ no longer provides a direct characterisation of tree-based networks in terms of matchings.

\begin{theorem}
\label{teo:f-s-s charact} 
Let $N=(V, A)$ be a phylogenetic network on $X$. Then the following are equivalent:
\begin{enumerate}[{\rm (i)}]
\item $N$ is tree-based.

\item $N$ has an antichain $A\subseteq V$ and a partition of $V$ into $|A|$ chains, each of which forms a path in $N$ ending at a leaf in $X$.

\item For all $U\subseteq V$, there exists a set of vertex disjoint paths in $N$, each ending at a leaf in $X$ such that each element of $U$ is on exactly one path.

\item The vertex set of $N$ can be partitioned into a set of vertex disjoint paths each of which ends at a leaf in $X$.

\item The bipartite graph $\mathcal{G}_N$ has a matching of size $|V|-|X|$.
\end{enumerate}
\end{theorem}

\subsection{Characterising temporal tree-based networks } 

The antichain-to-leaf property is not sufficient to characterise tree-based networks. However, Theorem~2.2 in~\cite{francis2016new} shows that this property is sufficient to characterised binary temporal tree-based networks. In this section, we generalise this result to arbitrary temporal tree-based networks. The proof of this generalisation uses a different approach to that taken in establishing Theorem~2.2 in~\cite{francis2016new}; the latter relies on the equivalence of (i) and (iii) in Theorem~\ref{teo:zhang} which, as discussed in Section~\ref{sec:bip_tree}, does not generalise to arbitrary tree-based networks.

We say that a phylogenetic network $N$ is \emph{temporal} if there exists a map $\lambda: V\rightarrow \mathbb R$, called a \emph{temporal map for $N$}, such that $\lambda(u) < \lambda(v)$ if $(u, v)$ is a tree arc, and $\lambda(u)=\lambda(v)$ if $(u, v)$ is a reticulation arc. In other words, if $N$ is temporal, then a `time stamp' can be assigned to each vertex such that time increases along tree arcs and remains constant along reticulation arcs.

\begin{theorem}
\label{temporalthm}
Let $N$ be a temporal network. Then $N$ is tree-based if and only if $N$ satisfies the antichain-to-leaf property. 
\end{theorem}

\noindent {\em Proof:}
If $N$ is tree-based, then it follows from the equivalence of (i) and (iii) in Theorem~\ref{teo:f-s-s charact} that $N$ satisfies the antichain-to-leaf property. 

We now establish the converse statement: If $N$ is a temporal network that satisfies the antichain-to-leaf property, then $N$ is tree-based.  We use induction on the 
number $n$ of vertices in $N$. If $n=1$, then $N$ consists of a single vertex, so $N$ is tree-based. Now assume that $n\ge 2$ and that every temporal network  with at most $n-1$ vertices and satisfying the antichain-to-leaf property is tree-based.
Let $\lambda$ be a temporal mapping for $N$. Let $\rho$ denote the root of $N$ and let $U=\{u_1, u_2, \ldots, u_k\}$ denote the set of children of $\rho$.

We first show that no vertex in $U$ is a reticulation. To see this, assume that $U$ contains a reticulation $u$. Since $u$ is a reticulation, $\lambda(u)=\lambda(\rho)$. Furthermore, as $N$ has no parallel arcs, $u$ has a parent $p_1$ that is not $\rho$. Since
$$\lambda(p_1)=\lambda(u)=\lambda(\rho),$$
it follows that $p_1$ is not a tree vertex. Therefore $p_1$ is a reticulation with at least one parent, $p_2$ say, that is not $\rho$ and $\lambda(p_2)=\lambda(p_1)$, in particular, $p_2$ is also a reticulation. It follows that $p_2$ has a parent that is a reticulation. Since $N$ is acyclic, this process has to eventually terminate. But the only way for this to happen is that there is reticulation in $U$ whose only parent is $\rho$; a contradiction. Thus each element in $U$ is a tree vertex. In particular, it follows that $U$ and every subset of $U$ is an antichain.

If $U$ contains a leaf $u$, let $N'$ denote the phylogenetic network obtained from $N$ by deleting $u$ and its incident arc. Then, as $N$ is temporal and has the antichain-to-leaf property, it immediately follows that $N'$ is temporal and has the antichain-to-leaf property. Therefore, by the induction assumption, $N'$ is tree-based as it has $n-1$ vertices. In turn, this implies that $N$ is tree-based. Thus we may assume that every vertex in $U$ is a tree vertex.

Without loss of generality, we may assume that $U_1=\{u_1, u_2, \ldots, u_j\}$ is the set of vertices in $U$ such that
$$\lambda(\rho) < \lambda(u_1) = \lambda(u_2) = \cdots = \lambda(u_j) < \lambda(v)$$
for all $v\in V(N)-\{\rho, u_1, u_2, \ldots, u_j\}$.  Let $N'$ be the rooted acyclic digraph obtained from $N$ by first contracting each of the arcs $(\rho, u_i)$, where $i\in \{1, 2, \ldots, j\}$, and then repeatedly deleting an arc from each non-trivial parallel class and suppressing any resulting non-root degree-two vertex. We next show that $N'$ is a temporal network satisfying the antichain-to-leaf property. We begin by determining which arcs of $N$ are deleted when obtaining $N'$ from $N$.

First observe that if an arc $e$ of $N$ is deleted in the process of obtaining $N'$ from $N$, then $e$ is a reticulation arc. Moreover, under the temporal mapping $\lambda$, the end vertices of $e$ are each assigned the value $\lambda(u_1)$. Let $w$ be a reticulation of $N$ such that $\lambda(w)=\lambda(u_1)$. If $w$ has two (distinct) reticulation parents, then $N$ does not satisfy the antichain-to-leaf property, and so at most one parent of $w$ is a reticulation and all of its remaining parents are in $\{u_1, u_2, \ldots, u_j\}$. This implies that $w$ has a unique ancestor, possibly itself, that is a reticulation in which every parent is in $\{u_1, u_2, \ldots, u_j\}$. It is now easily seen that, up to choosing which arcs to delete from a non-trivial parallel class, the only possible arcs of $N$ that are deleted when obtaining $N'$ from $N$ are those arcs of the form $(u_i, w)$, where $i\in \{1, 2, \ldots, j\}$ and $w$ is a reticulation in which $\lambda(w)=\lambda(u_1)$. In particular, if $w$ has a reticulation parent, then all arcs of the form $(u_i, w)$ are deleted, while if $w$ has no reticulation parent, that is, all its parents are in $U_1$, then all arcs except one of the from $(u_i, w)$ are deleted.

Let $M$ denote the rooted acyclic digraph obtained from $N$ by deleting the arcs determined in the last paragraph. Note that $N'$ is obtained from $M$ by contracting $(\rho, u_i)$ for all $i\in \{1, 2, \ldots, j\}$ and suppressing all non-root degree-two vertices. From this viewpoint, it is now easily seen that $N'$ is indeed a phylogenetic network. Furthermore, the mapping $\lambda'$ of the vertices of $N'$ to $\mathbb R$ inherited by $\lambda$ is a temporal mapping of the vertices of $N'$, so $N'$ is temporal. Lastly, to see that $N'$ satisfies the antichain-to-leaf property, observe that a subset of vertices of $N'$ is an antichain of $N'$ if and only if it is an antichain of $N$. Therefore, by again considering $M$, as $N$ satisfies the antichain-to-leaf property, $N'$ satisfies the antichain-to-leaf property. Hence, by induction, $N'$ is tree-based as it has at least one less vertex than $N$.

Let $T'$ be a base tree of $N'$. We complete the proof by extending $T'$ to a base tree of $N$. Let $S'$ be a support tree of $T'$ in $N'$, and note that $S'$ can be seen as an embedding of $T'$ in $N'$. View the arcs of $S'$ as arcs in $N'$ in the obvious way so that if $(\rho', v)$ is an arc in $S'$, where $\rho'$ is the root of $N'$ and $v\not\in U_1$, then the corresponding arc in $N$ is the unique arc directed into $v$. Denote this set of arcs in $N$ as $A'$. Let $O_1$ denote the subset of $U_1$ whose children are all reticulations, that is, $O_1$ is the subset of omnians in $U_1$. Then, in $N$, the vertices not incident with an edge in $A'$ are the vertices in $O_1$, all non-root degree-two vertices that were suppressed in constructing $N'$, and possibly the root $\rho$.

Let $V_1$ be the set of reticulations in $N$ whose parents are entirely in $U_1$. If $o$ is a vertex in $O_1$ with no child in $V_1$, then $V_1\cup \{o\}$ is an antichain. However, every child of $o$ is a reticulation with an ancestor in $V_1$. It follows that $V_1\cup \{o\}$ does not satisfy the antichain-to-leaf property. Thus if $o$ is a vertex in $O_1$, then it has at least one child in $V_1$. We now extend $A'$ to the set of arcs of support tree for $N$.

Since $O_1$ is an antichain, there is a set $P$ of disjoint paths with $|O_1|=|P|$ such that, for each $o\in W_1$, there is a path in $P$ starting at $o$, ending at a tree vertex, and for which each intermediate vertex is a reticulation. To see that we may assume that the second vertex in each of the paths in $P$ is a vertex in $V_1$, suppose that this assumption is not possible. Then, by Hall's Theorem~\citep{hall}, there are subsets $O'_1$ of $O_1$ and $V'_1$ of $V_1$ such that
$$N(O'_1)\cap V_1=V'_1$$
and
$$|O'_1| < |V'_1|,$$
where $N(O'_1)$ denotes the set of outgoing neighbours of $O'_1$. It now follows that, as each reticulation $w$ with $\lambda(w)=\lambda(u_1)$ has a unique ancestor in $V_1$, the union of $V_1-V'_1$ and $O'_1$ is an antichain but it does not satisfy the antichain-to-leaf property; a contradiction. With this in hand, extend $A'$ by taking the union of $A'$ and the following sets of arcs:
\begin{enumerate}[(i)]
\item $\{(\rho, u_i): i\in \{1, 2, \ldots, j\}\}$;

\item the union of the arcs in $P$; and

\item for each $v\in V_1$ not on a path in $P$, the union of the arcs in a path starting at a vertex in $O_1$ whose second vertex is $v$, ending at a tree vertex, and for which each intermediate vertex is a reticulation.
\end{enumerate}
It is easily checked that the resulting extension of $A'$ is the set of arcs of a support tree for $N$. This completes the proof of the theorem. \qed

\bigskip

\begin{figure}[ht]
\centering
\includegraphics[scale=1.0]{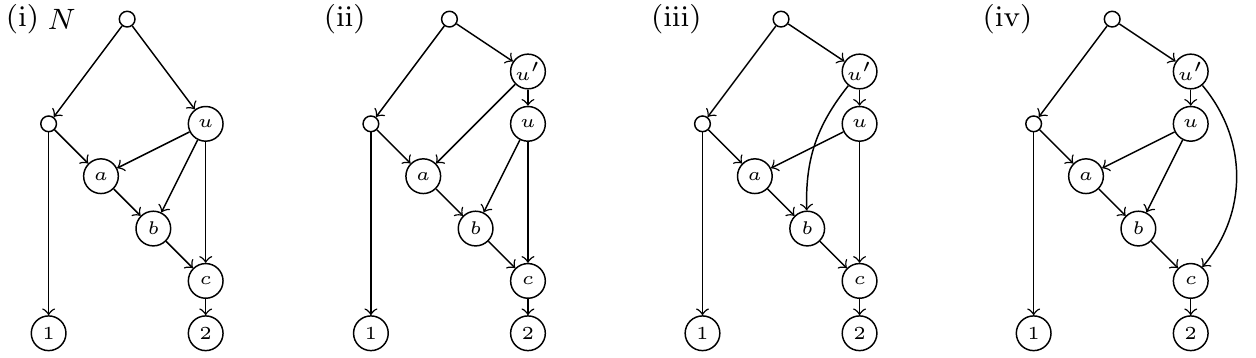}
\caption{(i) A temporal network $N$ and (ii)--(iv) its binary refinements none of which are temporal.}
\label{fig:no-tempo}
\end{figure}

\noindent {\em Remark:} 
A nonbinary  temporal network need not have a binary temporal refinement. For example, the nonbinary tree-based network shown in Fig.~\ref{fig:no-tempo}(i) is temporal, but none of its three binary refinements are temporal.   This means that one cannot establish the non-trivial (`if') direction of Theorem~\ref{temporalthm} by simply applying the corresponding result for binary networks  from \cite{francis2016new} 
to the nonbinary setting (noting that if a nonbinary network satisfies the antichain-to-leaf property then any binary refinement of it does also).

\subsection{Deviation measures}

The notion of tree-based is an all-or-nothing approach to formalising the extent to which a phylogenetic network has an underlying phylogenetic tree. Yet, 
certain nontree-based phylogenetic networks are nevertheless `close' to being tree-based. For example, if we adjoin a single new leaf, $5$ say, to the phylogenetic network $N$ shown in Fig.~\ref{fig:net2}(i) by subdividing the arc $(x, z)$ with a new vertex $u$ and adding the arc $(u, 5)$, then we construct a tree-based network $N'$. Motivated by such examples, \cite{francis2016new} considered three efficiently computable indices each measuring the closeness of a binary phylogenetic network to being tree-based. In this section, we interpret these indices for arbitrary phylogenetic networks.

Let $N=(V, A)$ be a phylogenetic network on $X$. The operation of adjoining a new leaf $y$ to $N$ by subdividing an arc of $N$ with a new vertex $u$ and adding the arc $(u, y)$ is called {\em attaching a new leaf to $N$}. Note that $u$ is a tree vertex in the resulting phylogenetic network. The three indices are as follows:
\begin{enumerate}[(I)]
\item The minimum number $l(N)$ of leaves in $V-X$ that must be present as leaves in a rooted spanning tree of $N$.

\item The minimum number $p(N)=d(N)-|X|$, where $d(N)$ is the smallest number of vertex disjoint paths that partition the vertices of $N$.

\item The minimum number $t(N)$ of leaves that need to be attached to $N$ so the resulting network is tree-based.
\end{enumerate}

Each of these measures is well-defined, non-negative and equal to zero if and only if $N$ is tree-based. For (I), this relies on Lemma~\ref{lemeqz}, while for (II), we consider the equivalence of (i) and (iv) in Theorem~\ref{teo:f-s-s charact}. To see that (III) is well defined, we can proceed by attaching a new leaf to each reticulation arc in $N$. In particular, this means that the resulting phylogenetic network $N'$ has no omnians, and therefore the bipartite graph $\mathcal{B}_{N'}$ is empty of edges, in which case $N$ is (trivially) tree-based by applying Theorem~\ref{teo:iersel_charact}. Note that, since (iii) implies (i) in Theorem~\ref{teo:iersel_charact}, it is enough to attach a new leaf to one reticulation arc for each maximal path that starts and ends with a vertex in $O$ in $\mathcal{B}_N$ instead of making this attachment to every reticulation arc in $N$. This way we `break' these unwanted paths in $\mathcal{B}_N$ to produce a tree-based network.

Surprisingly, all three indices are identical for binary phylogenetic networks~\citep{francis2016new} and, as it turns out, for arbitrary phylogenetic networks. Moreover, as in~\cite{francis2016new}, the measures are computable in polynomial-time in the size of $N$ as they can be written in terms of the size of maximum matching in the bipartite graph $\mathcal G_N$. The proof of the next theorem is essentially the same as the analogous result in~\cite{francis2016new} and is omitted.

\begin{theorem}
\label{theo}
Let $N=(V, A)$ be a phylogenetic network on $X$. Then
$$l(N)=p(N)=t(N)=(|V|-|X|)-m(\mathcal{G}_N),$$
where $m(\mathcal{G}_N)$ is the size of a maximum matching of $\mathcal{G}_N$.
\end{theorem}

For a phylogenetic network $N$, the following theorem expresses the three measures in terms of maximum-sized matchings in $\mathcal B_N$, thus providing an alternative viewpoint.

\begin{theorem}
\label{omni}
Let $N$ be a phylogenetic network, and let $\{O, R\}$ be the vertex bipartition of $\mathcal B_N$, where $O$ and $R$ are the sets of omnians and reticulations in $N$. Then
$$l(N)=p(N)=t(N)=|O|-m(\mathcal B_N),$$
where $m(\mathcal B_N)$ is the size of a maximum matching of $\mathcal B_N$.
\end{theorem}

\noindent {\em Proof:}
Let $M$ be a maximum-sized matching of $\mathcal B_N$, and let $O_u$  denote the subset of vertices in $O$ unmatched by $M$. By Theorem~\ref{theo}, it suffices to show that $t(N)=|O_u|$. We first establish $t(N)\leq |O_u|$. The proof is by induction on the size of $|O_u|$. If $|O_u|=0$, then, by Theorem~\ref{teo:iersel_charact}, $N$ is tree-based, so $t(N)=0$ and the inequality holds. Now assume that $|O_u|\geq 1$ and the inequality holds for all phylogenetic networks $N'$ with the property that, in relation to a maximum-sized matching of $\mathcal B_{N'}$, the number of unmatched vertices in $O'$ in $\mathcal B_{N'}$, where $O'$ is the set of omnians of $N'$, is at most $|O_u|-1$.

Let $u\in O_u$ and let $r$ be a child of $u$ in $N$. Note that $r\in R$. Let $N'$ be the phylogenetic network obtained from $N$ by attaching a new leaf $y$ to $(u, r)$. Let $t$ denote the parent of $y$. Since neither $u$ nor $t$ are omnians in $N'$, the bipartite graph $\mathcal B_{N'}$ can be obtained from $\mathcal B_N$ by deleting $u$ and its incident arcs. Since $u$ is unmatched in $\mathcal B_{N}$, it follows that $M$ is a matching in $\mathcal B_{N'}$. Moreover, as $M$ is a maximum-sized matching of $\mathcal B_N$, it is a maximum-sized matching of $\mathcal B_{N'}$. The number of unmatched omnians in $\mathcal B_{N'}$ is $|O_u|-1$ and so, by the induction assumption, $t(N')\leq |O_u|-1$. In particular, as $t(N)\leq t(N')+1$, we have $t(N)\leq |O_u|$.

We next establish the inequality $t(N)\geq |O_u|$. The proof is by induction on $t(N)$. If $t(N)=0$, then $N$ is tree-based and so, by Theorem~\ref{teo:iersel_charact}, $|O_u|=0$. In particular, the inequality holds. Suppose that $t(N)\geq 1$ and the inequality holds for all phylogenetic networks $N'$ with $t(N') \leq t(N)-1$.

Since $t(N)\geq 1$, we can adjoin a leaf $y$ to $N$ by subdividing an arc $(u,v)$ with a new vertex $t$ to obtain a phylogenetic network $N'$ with $t(N')=t(N)-1$. Note that this attachment preserves reticulations but may reduce the number of omnians by one. Let $M'$ be a maximum-sized matching of $\mathcal{B}_{N'}$ and let $O'_u$ be the subset of unmatched vertices in $O'$ in $\mathcal B_{N'}$. By the induction assumption, $t(N')\geq |O'_u|$.

It is easily checked that $\mathcal B_N$ can be obtained from $\mathcal{B}_{N'}$ as follows. If $u$ is not an omnian of $N$, then $\mathcal B_N$ and $\mathcal{B}_{N'}$ are identical. Otherwise, add the vertex $u$ to the set $O'$ and add $(u, v), (u, w_1), (u, w_2), \ldots, (u, w_k)$ to the set of arcs of $\mathcal B_{N'}$, where $v, w_1, w_2, \ldots, w_k$ are the children of $u$, to obtain $\mathcal B_N$. Now, consider the maximum-sized matching $M'$ of $\mathcal{B}_{N'}$. By construction, a maximum-sized matching of $\mathcal B_N$ has size at least $|M'|$. Thus
$$t(N)=t(N')+1\geq |O'_u|+1\geq |O_u|,$$
and so $t(N)\geq |O_u|$. This completes the proof of the theorem. \qed

\section{Embedded support trees}
\label{part2}

Let $N$ be a tree-based network on $X$ and let $S$ be an embedding in $N$ of a phylogenetic $X$-tree displayed by $N$. Then $S$ is a support tree for $N$ precisely if every vertex of $N$ is a vertex of $S$. Note that not every embedding in $N$ of a phylogenetic $X$-tree is a support tree for $N$. Moreover, a fixed base tree for $N$ may have at least two distinct support trees that corresponds to it. Denoting the set of support trees for $N$ by $Sup(N)$, the goal of this section is to enumerate the size of $Sup(N)$, that is, determine $|Sup(N)|$. One motivation for addressing this question is that counting the number of displayed trees in a general binary phylogenetic network is known to be \#P-complete \citep{linz}, and this holds even when the networks constrained to be temporal and tree-based.  However, for binary tree-based networks our results below show that counting support trees can be carried out polynomial-time. Note that counting support trees is different from counting displayed trees. The number of support trees a network has also provides a further measure of its `complexity':  a network with many support trees allows numerous possible evolutionary scenarios that combine `tree-like' evolution with reticulation events.

\begin{figure}[ht]
\center
 \includegraphics[scale=1.0]{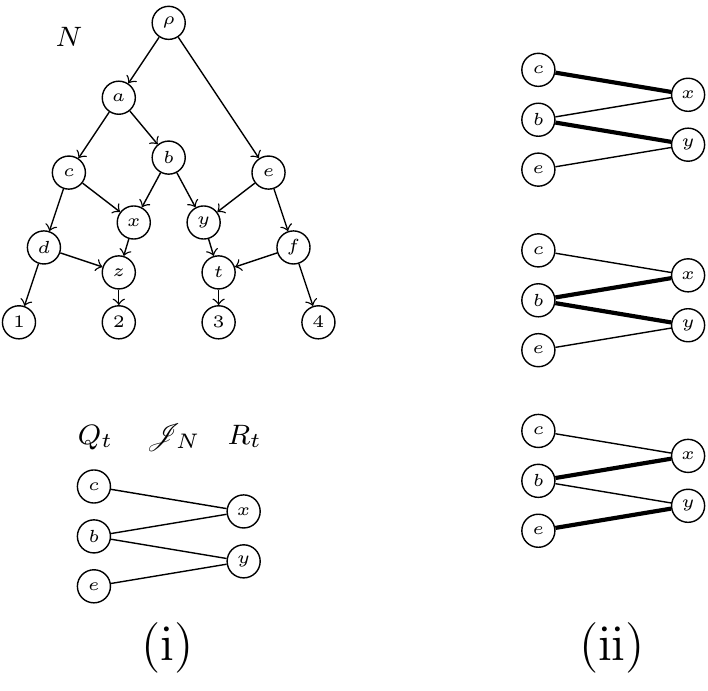}
 \caption{(i) A tree-based network $N$ and its bipartite graph $\mathcal{J}_N$ where the vertex $b$ is the only omnian in $Q_t$. (ii) The three supporting sets (in bold) for $\mathcal{J}_N$.}
 \label{fig:fromGtoJ}
\end{figure}

We begin by characterising the set of arcs of a support tree for $N$.  Let $N=(V, A)$ be a tree-based network. Let $R_t$ be the set of reticulations of $N$ with no reticulation parent, and let $Q_t$ be the set of vertices of $N$ with a child in $R_t$. Note that $Q_t$ and $R_t$ are disjoint as $Q_t$ consists of tree vertices. Let $\mathcal J_N$ be the bipartite graph with vertex partition $\{Q_t, R_t\}$ and arc set
$$\{\{q, r\}: q\in Q_t, r\in R_t, (q, r)\in A\}.$$
To illustrate, consider the phylogenetic network $N$ and the bipartite graph $\mathcal J_N$ shown in Fig.~\ref{fig:fromGtoJ}(i).

Let $N=(V, A)$ be a tree-based network. We say that a subset $E$ of edges of $\mathcal J_N$ is a {\em supporting set} if each omnian in $Q_t$ is incident with at least one edge in $E$ and each reticulation in $R_t$ is incident with exactly one edge in $E$. Note that if $E$ is a support set, then $E$ is not necessarily a matching. This is illustrated in Fig.~\ref{fig:fromGtoJ}(ii).

For a tree-based network $N$, an arc $(u, v)$ of $N$ is {\em arboreal} if either $u$ is a reticulation, or $v$ is a tree vertex or a leaf. Observe that if $S$ is a support tree for $N$, then the set of arboreal arcs of $N$ is a subset of the arcs of $S$. Also, if $(u, v)$ is arboreal, then $\{u, v\}$ is not an edge in $\mathcal J_N$. For the purposes of the next theorem and without ambiguity, we view each edge of $\mathcal J_N$ as the corresponding arc of $N$.

\begin{theorem}
Let $N=(V, A)$ be a tree-based network. Let $A'$ be a subset of $A$. Then $A'$ is the set of arcs of a support tree for $N$ if and only if $A'$ is the union of the set of arboreal arcs of $N$ and a supporting set for $\mathcal J_N$.
\label{valid}
\end{theorem}

\noindent {\em Proof:}
First suppose that $S=(V, A')$ is a support tree for $N$. Let $B$ be the subset of arcs in $A'$ that are not arboreal. To establish the necessary direction, it suffices to show that $B$ is a supporting set for $\mathcal J_N$. If $(u, v)\in B$, then, as $(u, v)$ is not arboreal, $u$ is a tree vertex and $v$ is a reticulation. Thus, $\{u, v\}$ is an edge of $\mathcal J_N$ unless there is a parent, $p$ say, of $v$ in $N$ that is a reticulation. But then $(p, v)$ is arboreal and so $(p, v)\in A'$ which, together with $(u, v)\in A'$, implies that $S$ is not a support tree for $N$; a contradiction. So $\{u, v\}$ is an edge of $\mathcal J_N$. Since $A'$ is the set of arcs of a support tree for $N$, if $r$ is a reticulation with no reticulation parent, then exactly one arc in $B$ is directed into $r$. Furthermore, if $q$ is an omnian tree vertex, then at least one arc in $B$ is directed out of $q$. It now follows that $B$ is a supporting set for $\mathcal J_N$.

Now suppose that $A'$ is the union of the set of arboreal arcs of $N$ and a supporting set $B$ for $\mathcal J_N$. To show that $A'$ is the set of arcs of a support tree for $N$, it suffices to show that, for every non-root vertex $v$ of $N$, there is exactly one arc in $A'$ directed into it and, if $v\not\in X$, at least one arc directed out of it. If $v$ is tree vertex or a leaf, then the unique arc directed into $v$ is arboreal, and so it is in $A'$ and there is exactly one such arc. Furthermore, if $v$ is a tree vertex and there is no arc in $A'$ directed out of it, then $v$ is an omnian. Also, as $N$ is tree-based, it is easily seen that at least one child of $v$ has the property that all of its parents are tree vertices. Since $B$ is a support set for $\mathcal J_N$, it follows that $B$ contains an arc directed out of $v$. Now assume that $v$ is a reticulation. If $v$ has reticulation parent, then, as $N$ is tree-based, it has exactly one reticulation parent, $p$ say, and $(p, v)$ is the unique arboreal arc directed into $v$. Furthermore, by definition, no arc in $B$ is directed into $v$. On the other hand, if $v$ has no reticulation parent, then it is a vertex in $\mathcal J_N$, in which case, there is exactly one arc in $B$ directed into $v$. As no arboreal arc is directed into $v$, it follows that $A'$ is the set of arcs of a support tree for $N$. This completes the proof of the theorem. \qed

A direct consequence of Theorem~\ref{valid} is the following corollary.

\begin{corollary}
\label{coro:valid_choice}
Let $N$ be a tree-based network. Then $|Sup(N)|$  is equal to the number of supporting sets for $\mathcal{J}_N$.
\end{corollary}

\begin{figure}[ht]
\centering
\includegraphics[scale=1.0]{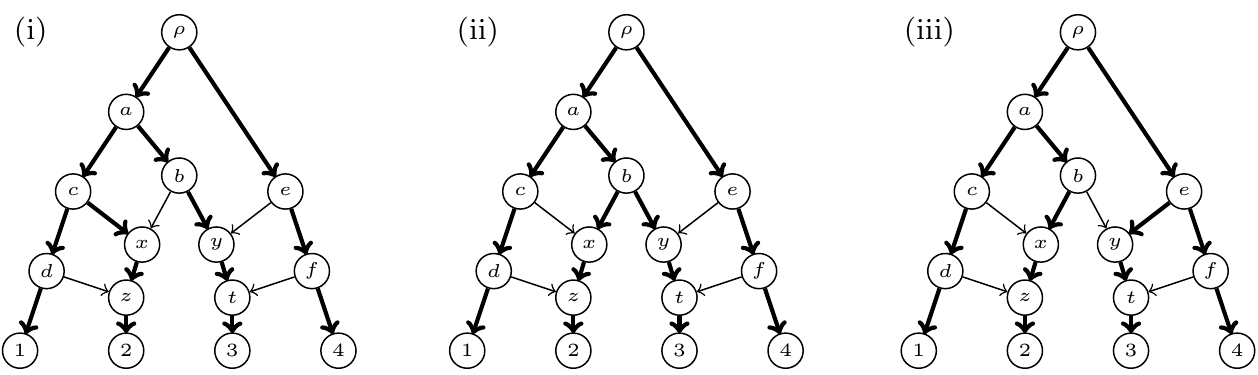}
 \caption{The set of support trees (arcs in bold) for the tree-based network shown in Fig.~\ref{fig:fromGtoJ}.  (i)--(iii) The support trees corresponding to the supporting sets $\{\{c,x\},\{b,y\}\}$, $\{\{b,x\},\{b,y\}\}$, and $\{\{b,x\},\{e,y\}\}$ of $\mathcal{J}_N$, respectively.}
 \label{fig:support_trees}
\end{figure}

For the phylogenetic network $N$ shown in Fig.~\ref{fig:fromGtoJ}(i), the three supporting sets for $\mathcal{J}_N$ are shown in Fig.~\ref{fig:fromGtoJ}(ii) while, in Fig.~\ref{fig:support_trees}, the three corresponding support trees (in bold) are shown.

For the remainder of this section, we will restrict attention to binary tree-based networks. \blue{For the purposes of the proof of the next lemma and the rest of this section, if a component of graph $G$ consists of a cycle (resp.\ a path), we refer to the component as a {\em cycle component} (resp.\ {\em path component}) of $G$.}

\begin{lemma}
Let $N$ be a binary tree-based network. Then each connected component of $\mathcal J_N$ is one of the following:
\begin{enumerate}[{\rm (i)}]
\item a \red{cycle},

\item a path whose end vertices are tree vertices in $N$ neither of which are omnians, and

\item a path whose end vertices are tree vertices in $N$ exactly one of which is an omnian.
\end{enumerate}
\label{components1}
\end{lemma}

\blue{
\begin{proof}
Since $N$ is binary, each vertex of $\mathcal J_N$ has degree at most two, and so each component of $\mathcal J_N$ is either a \red{cycle} or a path. Since each vertex of $\mathcal J_N$ corresponding to a reticulation has degree two, it follows that if $u$ is a degree-one vertex of $\mathcal J_N$, then $u$ is a tree vertex. Note that no vertex of $\mathcal J_N$ has degree zero.
Now let $P=u_1\, r_1\, u_2\, r_2\, \cdots\, r_{k-1}\, u_k$ be the maximal path of a path component of $\mathcal J_N$. For all $i$, the vertices $u_i$ correspond to tree vertices of $N$ and the vertices $r_i$ correspond to reticulations of $N$. Assume that both $u_1$ and $u_k$ are omnians in $N$. Let $r_0$ denote the reticulation child of $u_1$ that is not $r_1$ and let $r_k$ denote the reticulation child of $u_k$ that is not $r_{k-1}$ in $N$. Note that $r_0\neq r_k$; otherwise, $r_0$ is a vertex in $\mathcal J_N$ as both of its parents are tree vertices. Furthermore, by the maximality of $P$, neither $r_0$ or $r_{k-1}$ corresponds to a reticulation in $\mathcal J_N$. Let $u_0$ and $u_{k+1}$ denote the reticulation parents of $r_0$ and $r_k$, respectively, that are not $u_1$ and $u_k$ in $N$. Observe that $u_1$ and $u_{k+1}$ are omnians. Therefore each of $u_0, u_1, \ldots, u_{k+1}$ are omnians, and it follows that
$$u_0\, r_0\, u_1\, r_1\, \cdots\, u_k\, r_k\, u_{k+1}$$
corresponds to a maximal path in the bipartite graph $\mathcal B_N$ as defined in Section~\ref{part1}. As $N$ is binary and this maximal path begins and ends with omnians, it follows by Theorem~\ref{teo:iersel_charact} that $N$ is not tree-based; a contradiction. Thus at most one of $u_1$ and $u_k$ is an omnian. The lemma immediately follows.
\end{proof}
}

Using $\mathcal B_N$, Theorem~2.14 in~\cite{jetten2015characterising} gives the following upper bound for the number of base trees for a binary tree-based network $N$:
\begin{align}
\blue{|Sup(N)| \leq} 2^c\cdot \prod_{P\in \pi(\mathcal B_N)} {\textstyle \frac{1}{2}}(v(P)+3),
\label{eqn}
\end{align}
where $c$ is the number of \red{cycle} components in $\mathcal B_N$, $\pi(\mathcal B_N)$ is the set of path components in $\mathcal B_N$ with terminal vertices in $R$, and $v(P)$ is the number of vertices in $P$. \blue{In Theorem~\ref{count} we will provide an exact expression for $|Sup(N)|$ which turns out to be equivalent to the right-hand-side of (\ref{eqn}); thereby showing that
 (\ref{eqn}) is actually an equality. We do this by relating the connected components of $\mathcal B_N$ and $\mathcal J_N$ in Lemma~\ref{components2}.}

\begin{theorem}
Let $N$ be a binary tree-based network. Then
$$|Sup(N)|=2^c\cdot \prod_{P\in \pi(\mathcal J_N)} {\textstyle \frac{1}{2}}(v(P)+1),$$
where $c$ is the number of \red{cycle} components in $\mathcal J_N$, $\pi(\mathcal J_N)$ is the set of path components in $\mathcal J_N$ without an omnian terminal vertex, and $v(P)$ is the number of vertices in path component $P$.   \blue{In particular, $|Sup(N)|$ can be computed in time polynomial in the size of $N$.}
\label{count}
\end{theorem}

\noindent {\em Proof:}
By Corollary~\ref{coro:valid_choice}, we establish the theorem by showing that the number of supporting sets for $\mathcal J_N$ equates to
\begin{align}
2^c\cdot \prod_{P\in \pi(\mathcal J_N)} {\textstyle \frac{1}{2}}(v(P)+1).
\label{eqncount}
\end{align}
To do this, it is enough to independently consider how each component contributes to a supporting set for $\mathcal J_N$. Fixing a component, let $B$ be the subset of edges of the component contained in a supporting set. If the component is a \red{cycle}, then the tree vertices in the component are omnians, and it follows that there are exactly two choices for $B$. Now suppose that the component is a path
$$P=q_1\, r_1\, q_2\, r_2\, \cdots\, r_k\, q_{k+1}.$$
First assume that neither $q_1$ nor $q_{k+1}$ is an omnian. It is easily checked that $B$ contains the edge $\{q_1, r_1\}$ if and only if $B$ consists of the edges
$$\{q_1, r_1\}, \{q_2, r_2\}, \ldots, \{q_k, r_k\}.$$
An analogous conclusion holds if $B$ contains the edge $\{q_{k+1}, r_k\}$. Furthermore, it is easily checked that if $B$ contains neither $\{q_1, r_1\}$ nor $\{q_{k+1}, r_k\}$, then exactly one omnian in $P$ is incident with two edges in $B$. In particular, for each $i\in \{2, 3, \ldots, k\}$, the set $B$ contains the edges $\{q_i,r_{i-1}\}$ and $\{q_i, r_i\}$ if and only if $B$ consists of the edges
$$\{q_2, r_1\}, \{q_3, r_2\}, \ldots, \{q_i, r_{i-1}\}, \{q_i, r_i\}, \ldots, \{q_k, r_k\}.$$
Thus the number of possibilities for $B$ is the number of tree vertices in $P$, that is, $\frac{1}{2}(v(P)+1)$. Second assume exactly one of $q_1$ and $q_{k+1}$ is an omnian. Without loss of generality, we may assume $q_1$ is an omnian, in which case, $B$ must contain the edge $\{q_1, r_1\}$ and so there is precisely one choice for $B$. Multiplying the number of choices for each contributing component  gives~(\ref{eqncount}). 

\blue{To count $Sup(N)$ in time polynomial in the size of $N$ one simply constructs the graph $\mathcal J_N$, determines the number $c$ of its cycle components, and  the set
$\pi(\mathcal J_N)$ of its path components; then for each such path component $P$, counts the number of vertices $v(P)$ in $P$ and insert these quantities and $c$ 
into the  expression in Theorem~\ref{count}.}

\qed

\blue{As a simple application of Theorem~\ref{count} to a biological example, the network from \cite{marc}  involving three ancient hybridization events in the evolution of bread wheat (studied in \cite{francis2015tree}, Fig. 4) gives rise to the bipartite graph $\mathcal J_N$ that consists of three disjoint paths of length 3 (the midpoint of each path being a reticulation). For this example, $c=0$ and $v(P)=3$ for each of the three path components $P$ in $\pi(\mathcal J_N)$, and so Theorem~\ref{count} gives $|Sup(N)| = 2^0 \cdot  [\frac{1}{2}(3+1)]^3 = 8$. In this example, each base tree is embedded exactly once.} 

We end this section by establishing the connection between the connected components of $\mathcal{B}_N$ and $\mathcal{J}_N$ to prove that~(\ref{eqn}) counts the number of support trees for $N$.

\begin{lemma}
Let $N$ be a binary tree-based network, and let $\{O, R\}$ be the vertex bipartition of $\mathcal B_N$, where $O$ and $R$ are the sets of omnians and reticulations in $N$, respectively. Then the following hold:
\begin{enumerate}[{\rm (i)}]
\item A subset $C$ of arcs of $N$ is the set of edges of a \red{cycle} component of $\mathcal B_N$ if and only if $C$ is the set of edges of a \red{cycle} component of $\mathcal J_N$.

\item A subset $P$ of arcs of $N$ is the set of edges of a path component of $\mathcal B_N$ with terminal vertices $r_1$ and $r_k$ in $R$ if and only if $P\cup \{\{q, r_1\}, \{q', r_k\}\}$ is a path component of $\mathcal J_N$, where $q$ and $q'$ are non-omnian tree vertices in $N$.
\end{enumerate}
\label{components2}
\end{lemma}

\noindent {\em Proof:}
To see (i), observe that if $C$ is a \red{cycle} component of $\mathcal B_N$, then each vertex of $O$ in $C$ has out-degree two and so each such vertex is a tree vertex. In turn, this implies that each vertex of $R$ in $C$ has the property that each of its parents is a tree vertex. Part (i) is easily deduced from this observation.

For the proof of (ii), observe that if $P$ is a path component of $\mathcal B_N$ with terminal vertices $r_1$ and $r_k$ in $R$, then each vertex of $O$ in $P$ is a tree vertex and each vertex of $R$ in $P$ has the property that its parents are tree vertices. Note that, if a terminal vertex, $r_1$ say, does not have this property, then one of its parents, $p$ say, is a reticulation. But then $p$ is an omnian and so it is a vertex in $O$, contradicting the assumption that $P$ is a component. Using this observation, it is easily checked that (ii) holds. 

\qed

The next theorem immediately follows from Theorem~\ref{count} and Lemma~\ref{components2}.

\begin{theorem}
\label{dutchcount}
Let $N$ be a binary tree-based network, and let $\{O, R\}$ be the vertex bipartition of $\mathcal B_N$, where $O$ and $R$ are the sets of omnians and reticulations in $N$, respectively. Then
$$|Sup(N)|=2^c\cdot \prod_{P\in \pi(\mathcal B_N)} {\textstyle \frac{1}{2}}(v(P)+3),$$
where $c$ is the number of \red{cycle}s in $\mathcal{B}_N$, $\pi(\mathcal{B}_N)$ is the set path components in $\mathcal B_N$ with terminal vertices in $R$, and $v(P)$ is the number of vertices in $P$.
\end{theorem}

\section{Concluding comments}

In this paper, we have shown how recent characterisations and properties of tree-based networks (based on disjoint path conditions or matchings in bipartite graphs) as well as proximity measures, can be extended from binary phylogenetic networks to arbitrary phylogenetic networks. In some instances, the extensions are possible by adapting the approach used in the binary case. However, other results, for example, Theorem~\ref{temporalthm} concerning the antichain-to-leaf property characterisation of tree-based for temporal networks, seem to require a completely different approach.

In the second part of the paper, we investigated the problem of determining, for a given tree-based network $N$, the number of support trees for $N$. We introduced the bipartite graph $\mathcal{J}_N$ and showed that there is a one-to-one correspondence between the supporting sets for $\mathcal J_N$ and the support trees for $N$. \blue{We then restricted this focus to binary networks, and this enabled us to determine the number of support trees when} $N$ is binary. Two questions immediately arise. What is this number when $N$ is not necessarily binary, and how do we distinguish when two support trees for $N$ result in the same base tree so that, instead of counting support trees, we count the number of base trees of $N$?
\blue{We leave these questions for future work.}

\begin{acknowledgements}
We would like to thank Momoko Hayamizu for helpful comments concerning Theorem~\ref{count} and the two anonymous reviewers for further suggestions.
JCP thanks the Biomathematics Research Centre of the University of Canterbury (and especially MS and CS) for hosting his visit, which led to this collaboration. 
We thank the (former) Allan Wilson Centre \red{and the New Zealand Marsden Fund} for funding support for this project.
\end{acknowledgements}


\bibliographystyle{apalike}
\DeclareRobustCommand{\VAN}[3]{#3}
\bibliography{biblio_tesi}  

\section{Appendix: Proof of Theorem~\ref{teo:f-s-s charact}.}

We begin with a lemma, which extends Lemma 3.1 of \cite{francis2016new}.

\begin{lemma}
\label{lem:f-s-s lemma} 
Let $T$ be a subdivision of a rooted tree with vertex set $V$. Then the following property holds:
\begin{enumerate}[{\rm (P)}]
\item For any nonempty subset $U$ of $V$, there exists a set of vertex disjoint (directed) paths in $T$ each of which ends at a leaf of $T$ and each vertex in $U$ lies on exactly one path.
\end{enumerate}
\end{lemma}

\noindent {\em Proof:}
As in the proof of Lemma 3.1 from \cite{francis2016new}, we use induction on the number $n$ of vertices of $N$. The result holds trivially for $n=1$, so suppose
that $n \geq 2$ and that (P) holds for all subdivisions of a rooted tree with at most $n-1$ vertices. Let $U$ be an arbitrary subset of the vertices of $T$.  Since
$n\geq 2$, it follows that $T$ either has
\begin{enumerate}[(i)]
\item a leaf $x$ whose parent, $u$ say, has degree 2, or

\item  a vertex $v$ that is the parent of $k \geq 2$ leaves, $x_1, x_2, \ldots, x_k$ say.
\end{enumerate}
Case (i) is handled in the same way as in the proof of Lemma~3.1 from \cite{francis2016new} to justify the induction step. 
For Case (ii), let $T'$ be the subdivision of a rooted  tree obtained from $T$ by deleting $x_2, x_3, \ldots, x_k$ and the incident edges $(v, x_i)$ for all $i=2, 3, \ldots, k$. Note that $T'$ has $n-(k-1)$ vertices. Let $Y=U\cap \{x_1, x_2, \ldots, x_k\}$. If $Y$ is empty, then let $U'=U$. By induction there is a set of vertex disjoint paths in $T'$ each of which ends at a leaf of $T'$ and each vertex in $U'$ lies on exactly one path. This set of paths also work for $U$ in $T$. On the other hand, if $Y$ is nonempty, let $U'=U-Y$. By induction, there is a set of at most $|U'|=|U|-|Y|$ vertex disjoint paths in $T'$ each of which ends at a leaf of $T'$ and each vertex in $U'$ lies on exactly one path. Adding the set of $|Y|$ (trivial) paths each of which consists of a distinct element in $Y$, we obtain a set of vertex disjoint paths in $T$ each of which ends at a leaf of $T$ and each vertex in $U$ lies on exactly one path. This establishes the induction step in Case (ii), and thereby Lemma~\ref{lem:f-s-s lemma}. \qed

Returning to the proof of Theorem~\ref{teo:f-s-s charact}, we establish the following implications between the stated conditions on $N$. First, by applying Lemma~\ref{lem:f-s-s lemma}, the same proof of Theorem~2.1 in \cite{francis2016new} for the binary case is valid for arbitrary phylogenetic networks to verify the sequence of implications: (i) $\Rightarrow$ (ii) $\Rightarrow$ (iii) $\Rightarrow$ (iv). Thus, we prove in addition that (iv) $\Rightarrow$ (v) and (v) $\Rightarrow$ (i) which, together with the previous implications, shows that (i)--(v) are equivalent.

First we prove that (iv) implies (v). Suppose the vertex set of $N=(V,A)$ can be partitioned into a set ${\mathcal P}$ of disjoint paths $P_1, P_2, \ldots, P_k$ each of which ends at a leaf in $X$.  Thus $k= |\mathcal{P}|=|X|$. Let $v_i$  (respectively $x_i$) be the first (respectively, last) vertex in $P_i$, and let $V_1' = V-\{x_1, \ldots, x_k\}, V_2' = V-\{v_1, \ldots, v_k\}$.
 Recall that $\mathcal{G}_N$ is the bipartite graph with vertex bipartition $\{V_1, V_2\}$, where $V_1=V_2=V$, and edge set
 $$\{\{v_1,v_2\}: v_1\in V_1, v_2\in V_2, (v_1, v_2)\in A\}.$$
Then each arc in a path from $\mathcal{P}$  corresponds to an edge $\{u,v\}$, where $u \in V_1' \subseteq V_1$ and $v \in V_2' \subseteq V_2$.  Moreover, these edges are vertex-disjoint, and since there are $|V|-|X|$ such edges, we obtain a matching of $\mathcal{G}_N$ of size $|V|-|X|$, as claimed.

Finally,  we prove that (v) implies (i). Note that none of the leaf vertices in $V_1$ is matched in the matching of size $|V|-|X|$ in $\mathcal{G}_N$, say $M$. In particular, all omnians of $N$ are matched by $M$ and this provides a matching in $\mathcal{B}_N$ that covers all omnians. Therefore, by Theorem~\ref{teo:iersel_charact}, $N$ is tree-based. \qed

\end{document}